\documentclass[preprint,prd,twoside,nofootinbib,
aps,superscriptaddress,tightenlines,preprintnumbers]{revtex4}
\usepackage{amsmath}
\usepackage{amssymb}
\usepackage{graphicx}
\usepackage{bm}
\usepackage{epsfig}

\newcommand\bi{\begin{itemize}}
\newcommand\ei{\end{itemize}}
\newcommand\be{\begin{equation}}
\newcommand\ee{\end{equation}}
\def\Dsl{\,\raise.15ex \hbox{/}\mkern-12.8mu D}

\newcommand{\vect}[1]{\mathbf{#1}}
\def\Dsl{\,\raise.15ex \hbox{/}\mkern-12.8mu D}
\newcommand{\nslash}{n\!\!\! /}
\newcommand{\bnslash}{\bar n\!\!\! /}
\newcommand{\Aslash}{A\!\!\! /}
\newcommand{\Dslash}{D\!\!\!\!/\,}
\newcommand{\Pslash}{\mathcal{P}\!\!\!\!/}

\newcommand{\bwt}{\begin{widetext}}
\newcommand{\ewt}{\end{widetext}}

\newcommand{\abs}[1]{\left\lvert #1\right\rvert}

\newcommand{\bra}[1]{\left\langle #1\right\rvert}
\newcommand{\ket}[1]{\left\lvert #1\right\rangle}
\newcommand{\Lqcd}{\Lambda_{\text{QCD}}}
\newcommand{\bea}{\begin{eqnarray}}
\newcommand{\eea}{\end{eqnarray}}
\newcommand{\e}{\mathrm{e}}
\newcommand{\Cumulant}[1]{\left\langle\!\!\!\left\langle #1 \right\rangle\!\!\!\right\rangle}
\newcommand{\cumulant}[1]{\langle\!\langle #1 \rangle\!\rangle}
\newcommand{\SCETa}{$\text{SCET}_{\text{I}}$}
\newcommand{\SCETb}{$\text{SCET}_{\text{II}}$}

\DeclareMathOperator{\Tr}{Tr}

\begin{document}

\title{Momentum Flow Correlations from Event Shapes:\\
Factorized Soft Gluons and Soft-Collinear Effective Theory}

\author{Christopher Lee}
\affiliation{Institute for Nuclear Theory, University of Washington, Seattle, WA  98195-1550, USA}

\author{George Sterman}
\affiliation{C.N.\ Yang Institute for Theoretical Physics\\ Stony Brook University,
Stony Brook, NY 11794-3840, USA}

\date{November 6, 2006\ \vspace{1cm}}

\preprint{INT-PUB 06-17}
\preprint{YITP-SB-06-47}

\bigskip

\begin{abstract}
The distributions of two-jet event shapes contain 
information on hadronization in QCD.  Near the two-jet limit,
these distributions can be described by convolutions of
nonperturbative event shape functions with the same distributions calculated in
 resummed perturbation theory.  The shape functions, in turn,
are determined by correlations of momentum
flow operators with each other and with light-like Wilson lines,
which describe the coupling of soft, wide-angle radiation to jets.   We observe that 
leading power corrections to the mean values of event shapes
are determined by the correlation of  a single momentum
flow  operator with the relevant Wilson lines.  This generalizes arguments for the universality of leading
power corrections based on the low-scale behavior of the
running coupling or resummation.
We also show how a study of the angularity event shapes
can provide information on correlations involving
multiple momentum flow operators, giving a window to the 
system of QCD dynamics that underlies the variety of event shape functions.
In deriving these results, we review, develop and compare factorization 
techniques in
conventional perturbative QCD and soft-collinear effective theory (SCET).
We give special emphasis to the elimination of double counting of momentum regions in
these two formalisms.
\end{abstract}

\maketitle

\section{Introduction}

In perturbative quantum chromodynamics (QCD),
 scattering amplitudes computed with massless 
partons suffer from infrared and collinear singularities,
which, however, cancel in suitably-defined infrared safe
cross sections.
Event shapes, $e(N)$, are numbers associated with
final states, $N$, defined so that the semi-inclusive cross sections
with initial state $I$ at center-of-mass energy $Q$,
\bea
\frac{d\sigma}{de}(Q) = \frac{1}{2Q^2}\ \sum_N \abs{ M_{I\rightarrow N}}^2\, (2\pi)^4\delta^4(Q - p_N)\delta(e - e(N))\, ,
\eea
are infrared safe for leptonic initial states (annihilation), and are factorizable
into parton distributions and infrared safe short-distance functions
for hadronic initial states.  Infrared safety requires, in general, that the event shape
$e(N)$ be equal for states that differ only by a rearrangement of
collinear partons or by the emission or absorption of zero-momentum partons.

Event shapes have been the subject of extensive study,
 partly as tools to explore the evolution of QCD dynamics 
 from weak to strong coupling (see \cite{review} for a review). 
 This paper will compare event shapes as treated
 in full QCD using factorization theorems and in 
 the QCD effective theory, soft-collinear
 effective theory (SCET).  In the process, we will derive
 a number of new results, relying on our ability 
to express infrared dynamics in
 terms of specific matrix elements in the full and effective
 theories.
 
Although infrared safe, and therefore calculable self-consistently in
perturbation theory, 
event shape cross sections at all but the very
highest energies are not usually well-approximated by perturbative results alone.
The relationship between perturbative and nonperturbative 
contributions to fully inclusive cross sections ($e(N)=1$) was formulated
early in terms of QCD sum rules \cite{qcdsr} based on fixed order
in perturbation theory and the operator product expansion, and later in terms of its high-order behavior
\cite{irrenorm}.   Eventually, it was realized that a similar
analysis can be applied to many semi-inclusive cross sections
that are infrared safe.

The leading nonperturbative contributions to 
the mean values of many event shapes 
enter at the level of $1/Q$  in leptonic annihilation experiments,
with $Q$ the total center-of-mass energy.  This dependence was
predicted from the high-order behavior of QCD
perturbation theory at infrared scales,
using analyses based on the running of the coupling
 \cite{oneoverQ,KS95,DW1,DMW}.  
These analyses naturally led to the proposal that
coefficients of the $1/Q$ power are universal up to calculable 
factors.

Power corrections in event shapes can be measured
by comparing fixed-order perturbative predictions to data from a range of $Q$,
and $1/Q$ corrections are readily seen in such comparisons \cite{review}.
For the mean values of many event shapes, these corrections are 
indeed universal,
in a sense that we will review below. This is a striking prediction for
nonperturbative effects, based on an analysis of perturbation 
theory at infrared scales.   

Beyond mean values, the distributions of event shapes require, as described
below, convolutions of resummed perturbative cross sections with nonperturbative 
``event shape functions" \cite{K98,KS99}, a different one for each event shape.
The event shapes summarize the leading power corrections for all the moments
of the event shape in question.

The roles and determinations of event shape functions
are in some ways analogous to those of parton distribution functions.   Parton
distributions can be defined in terms of matrix elements in 
QCD, and these matrix elements have universality properties
based on factorization.  
As we shall review below, event shape functions may also be defined as
 matrix elements in QCD or SCET, which describe correlations between energy flow 
in different directions in the presence of lightlike color sources \cite{KS99}.
The universality properties of event shape functions, however, are certainly less well
understood.

These issues can be addressed in the terminology of
perturbative QCD, using techniques of factorization and resummation.
At the present time, however, a growing body of work  
in jet physics relies on the language of 
soft-collinear effective theory to study power corrections
\cite{BMW,BLMW} and perturbative resummations \cite{BS1,Trott}.
With this in mind, we will use event shapes in the two-jet limit of
leptonic annihilation to compare treatments
based on factorized cross sections in
perturbative QCD \cite{KS95,K98,KS99} and alternatively in SCET
\cite{BMW,BLMW}.   We hope that this dual description will
be helpful to some readers.  
Of course, even for this limited set of observables, 
a full comparison for both perturbative resummation and
nonperturbative power corrections would require
a lengthy discussion.   In this paper, we will concentrate on
power corrections implied by these closely-related formalisms.

Not surprisingly, we will find that the two approaches are equivalent 
at the level of leading power corrections and shape functions that
are associated with soft gluon emission.  In both cases,
infrared dynamics will be described by matrix elements involving Wilson lines.
In particular, we will see that the soft shape functions
of the factorization theorem can be defined to correspond directly
to the functions that describe ultrasoft gluon radiation in these observables. 

Once we have discussed the formalism relating event shapes 
to matrix elements of momentum flow operators, we will  derive
a number of results of phenomenological interest.
The universality relations between different
$1/Q$ corrections proposed 
in Refs.\ \cite{oneoverQ,KS95,DW1,DMW}
 for mean values of  event shapes can be derived from the field-theoretic shape functions,
without invoking the dominance of single soft gluon emission
or related assumptions 
on the universality of the infrared running coupling \cite{A001}.
In particular, we will derive the universality of  
leading power corrections to the average values of a large class of event shapes 
entirely from factorization and the boost-invariance of products of
the relevant light-like Wilson lines.

To specify the full event shape function for an arbitrary infrared safe observable,
it is in principle
necessary to know all the energy flow correlators.
Relations between event shape functions, however, have been conjectured
for a particular class of event shapes, the angularities \cite{BKS03,scaling1,scaling2},
which include the thrust and jet broadening.
A scaling relation for the angularity shape functions follows from 
the assumption of negligible correlations between jet hemispheres for soft radiation.  We
will use insight gained from comparing factorized QCD and SCET to
derive explicit relations between energy correlations and violations of the scaling rule,
providing a set of measurements that relate directly to the correlations.

Before going further,
 it is important to emphasize that the strength of 
nonperturbative power corrections to any observable depends generically 
on the definition of perturbation theory chosen for that observable.  
In particular, in observable-specific perturbative schemes for the coupling based on
the method of effective charges \cite{effcharge}, as reviewed in
\cite{M06}, the coefficients of power corrections to average values tend to
decrease markedly compared to perturbative expansions
in $\rm \overline{MS}$ definitions.    
This method incorporates
measurements of the observable
in question directly into the renormalization scheme.
It therefore builds more information into perturbative expansions for these observables
than is possible in conventional schemes.
Certainly a better understanding
of the relationships between 
process-specific and process-independent 
approaches to perturbation theory would be helpful.
In the discussion of this paper, 
however, we will assume that the strong coupling, $\alpha_s$,
is defined in a process-independent fashion.

Our discussion begins in
the following section with a description of the 
class of event shapes that we study, and
specifically defines the angularity event shape functions.  In Sec.\ III, we
review and relate factorization formalisms for event shapes in full QCD 
and SCET, exhibit the
matrix elements that determine the soft gluon dynamics and
define the event shape functions.  We explore this relationship
further in Sec.\ IV, with a comparison
of the elimination
of double counting through zero-bin subtraction
in SCET \cite{0bin} and 
eikonal subtractions in QCD. 
In Sec.\ V we introduce momentum flow operators and draw the consequences
of boost invariance 
for their correlations in soft functions.  We show
that power corrections associated with jet functions are subleading
for a large class of angularities.

In Section VI we apply the formalism of Sec.\ V first to the
average values of event shapes, demonstrating the
universality properties of these mean values.
We go on to treat the scaling properties of angularities
beyond their mean values, and we discuss the information
on momentum flow correlators that is implicit in possible violations
of the scaling rule proposed in Refs.\ \cite{scaling1, scaling2}.

\section{Two-jet Event Shapes}

In this paper, we study event shapes that can be expressed in the form,
\begin{equation}
e = \frac{1}{Q}\sum_{i\in N}\abs{\vect{p}_i^\perp} f_e(\eta_i)\, ,
\label{eventshape}
\end{equation}
where the sum is over final state particles, and the transverse momenta 
and (pseudo-\!\!~)rapidities $\eta_i = \ln\cot(\theta_i/2)$ are 
measured with respect to the thrust axis\footnote{The thrust axis is the 
choice of axis that minimizes the quantity 
on the right-hand side of 
$\tau_a$ in Eq.~(\ref{shapedefs}) 
for $a=0$.}.
 Two examples of these are the $C$-parameter and angularities (which include the thrust), 
which are expressed in terms of the rapidities as \cite{Salam01,scaling1} 
\begin{equation}
C = \frac{1}{Q}\sum_{i\in N}\frac{3\abs{\vect{p}_i^\perp}}{\cosh\eta_i}, \quad
\tau_a = \frac{1}{Q}\sum_{i\in N}\abs{\vect{p}_i^\perp} \e^{-\abs{\eta_i}(1-a)}.
\label{shapedefs}
\end{equation}
In the limit $e\rightarrow 0$, the cross sections for all of these event shapes 
are dominated by
final states consisting of two perfectly-collimated jets.
For this reason, we refer to them as two-jet event shapes, and the limit
$e\rightarrow 0$ as the two-jet limit.  Power corrections to 
the distributions for these
event shapes enter at the level of $1/(eQ)^n$, starting at $n=1$,
in addition to corrections suppressed by additional powers of $Q$ 
\cite{oneoverQ,KS95,DW1,DMW,BMW,BLMW}.

Near $e=0$, all such event shapes generate double logarithms
in perturbation theory,
which in many cases can be resummed to next-to-leading
logarithms \cite{resumdistr,broadening1,CLS97} and beyond \cite{BKS03}.  
A quantitative description of these distributions, however, requires
nonperturbative input, which can be summarized in an {\it event shape function}, 
$S_{NP,e}$.
The physical cross section is then given as a convolution
of this shape function with the resummed perturbative function \cite{K98,KS99}, 
\bea
\frac{d\sigma(e,Q)}{de} =  \int_0 d\zeta\ S_{NP,e}(\zeta Q,\Lambda)\ 
\frac{d\sigma_{\rm PT}(e-\zeta,Q,\Lambda)}{de}\, .
\label{convol}
\eea
The mass $\Lambda$,
which we may think of as a scale comparable to, but larger 
than, $\Lambda_{\rm QCD}$, represents the 
boundary between perturbative and nonperturbative contributions.
Below, it will appear as the renormalization scale for certain matrix 
elements that define the nonperturbative function.

The nonperturbative function $S_{NP,e}(\zeta Q,\Lambda)$
for  event shape $e$ is independent
of the overall momentum scale, $Q$.    
Estimating $S_{NP,e}(\zeta Q,\Lambda)$ in Eq.~(\ref{convol})
from the plentiful data at the Z pole, for example, allows predictions of cross sections
at lower and higher energies.  These predictions successfully
describe $\rm e^+e^-$ data for the thrust and jet mass
distributions  for data over a wide range of center-of-mass
energies \cite{K98,KS99,GR02}.   

Our discussion below is focused on values of $e$
that describe 
two-jet events, which dominate final states
in leptonic annihilation.    Extensive data on
these shapes have been recorded, which may in principle
be mined for information on the process of hadronization
in QCD.  Extensions to multijet events are possible,
following Refs.\ \cite{BMDZ00,BSZ04}.  Certainly, the reasoning
below will require further development for these cases.

\section{Factorization  in the Two-Jet Limit}
\label{3a}

In this section, we review basic results on the factorization
of cross sections in the two-jet limit.  We begin with an outline of the 
factorization analysis in full QCD, 
in the notation of Ref.~\cite{BKS03},
followed by a discussion based
on soft-collinear effective theory \cite{BMW,BLMW}.   
Our discussion will apply to both perturbative and
nonperturbative contributions.

Most of the results of this section have been
given previously elsewhere,
 but we believe that a side-by-side presentation may help
to shed light on both formalisms. 
In particular, a comparison of the formalisms
will suggest the importance in both cases of the elimination of double counting.
This will be the subject of Sec.\ IV,
where we make use of the SCET discussion 
of ``zero-bin subtractions" in Ref.~\cite{0bin}.

\subsection{QCD matrix elements}

In the two-jet limit, the differential cross section (distribution) 
for a two-jet event shape $e$ factorizes into
a convolution of functions that characterize
the jets with a ``soft" function that describes wide-angle gluon emission \cite{BKS03},
\begin{equation}
\frac{d\sigma}{de} = \sigma_0(Q)\ \int de_n\, de_{\bar n}\, de_S\
\delta(e - e_n-e_{\bar n}-e_S) 
  J_n(Q,e_n)\, J_{\bar n}(Q,e_{\bar n})\, S_{n\bar{n}}(e_sQ) + {\cal O}(e^0)
\, ,
\label{firstfact}
\end{equation}
where $\sigma_0$ carries the overall dimensions, and can be defined
as the Born cross section to lowest order in $\alpha_s$.  In perturbation theory
this cross section behaves as the order $1/e$ 
times logarithms as $e$ vanishes, with contributions
at order $e^0$ from wide-angle three-jet events.

In Ref.~\cite{BKS03}, the perturbative resummation of the angularities
was studied, with the functions in (\ref{firstfact}) defined  in terms of QCD matrix elements.
We may start by defining 
 a set of path-ordered exponentials  or Wilson lines,
\bea
\Phi^{\rm (f)}_{\xi_c} (z)  =  P \exp\left[i g \int^{0}_{-\infty} d
\lambda\; \xi_c \cdot {\mathcal{A}}^{\rm (f)} (\lambda \xi_c+z )\right]\, ,
\label{patho}
\eea
where f labels a color representation.  The vector $\xi_c$,
which defines the path of the ordering, was taken in Ref.~\cite{BKS03} to
be off the light-cone, at least to start.

Following Ref.~\cite{BKS03},
the jet functions of (\ref{firstfact}) are defined in terms
of matrix elements
\begin{widetext}
\bea
\bar{J}_c^\mu (Q,e_{J_c})
&=&
  \frac{2}{Q^2}\, \frac{(2\pi)^6}{N_C} \; \sum\limits_{N_{J_c}}
{\rm Tr} \left[\gamma^\mu
\left<0 \left|
\Phi^{(\rm q)\dag}_{\xi_c}{}(0) q(0) \right| N_{J_c} \right>
  \left< N_{J_c}\left| \bar{q}(0) \Phi^{(\rm q)}_{\xi_c}(0)
\right| 0 \right>\right] \nonumber \\
& &\, \hspace{15mm} \times \, \delta(e_{J_c}-\bar
e(N_{J_c}))\,
\delta\left(Q - \omega(N_{J_c})\right)\,
\delta^2(\hat n_{J_c} - \hat n(N_{J_c}))\, ,
\label{jetdef}
\eea
\end{widetext}
where $c=n,\bar{n}$ labels the direction of the jet as above.
The jet functions are constructed from the squared amplitudes for the 
quark (or other partonic) field to produce states $N_{J_c}$
with total energy $Q$, with a momentum whose direction $\hat{n}$
is in a fixed direction 
$\hat{n}_{J_c}$, and with 
a fixed contribution $e_{J_c}\ll 1$ to the event shape in question.
The amplitudes are rendered gauge invariant by multiplying 
the partonic fields by ordered exponentials (\ref{patho})
in the $\xi_c$ directions and in the quark representation.

To define the soft function, we introduce ``eikonal" cross sections,
\bea
\bar{\sigma}^{(\mbox{\tiny  eik})}\left(\mu, e\right)\!\!\!
& \equiv & \frac{1}{N_C}\
\sum_{N_{\mbox{\tiny eik}}} \left< 0
\left| \Phi^{(\bar {\rm q})}_{\bar{n}}{}^\dagger(0)
\Phi^{(\rm q)}_n{}^\dagger(0)
\right| {N_{\mbox{\tiny eik}}} \right>
\nonumber \\
& \ & \times\;
\left< N_{\mbox{\tiny eik}}
\left| \Phi^{(\rm q)}_n(0)
\Phi_{\bar{n}}^{(\bar{\rm q})}(0)
\right| 0 \right>  \ \delta\left(e - e(N_{\mbox{\tiny eik}}) \right)
\, ,
\label{eikdef}
\eea
in which final states $N_{\mbox{\tiny eik}}$
are produced by products of Wilson lines
in directions $n^\mu$ and $\bar{n}^\mu$,
which we may take to be opposite-moving.
The eikonal cross section, Eq.~(\ref{eikdef}) 
must be renormalized, with scale $\mu$.
It provides a good
approximation for soft radiation that is not collinear
to these vectors, but it also contains collinear-singular
radiation parallel to the directions of the lines.
The collinear enhancements are already taken into
account in the jet function, and the soft function
$S$ in Eq.~(\ref{firstfact}) must be
defined in such a way as not to double-count
these regions.  In fact, it is easier than it
might seem to avoid double-counting.
This is because we can apply the same factorization
to the eikonal as to the full cross section, factoring
the {\it same} soft function $S$ from a set of ``eikonal"
jet functions, which can themselves be defined in
terms of matrix elements as
\bea
\bar{J}_c^{(\mbox{\tiny eik})}\left(Q,e_c \right)
& \equiv  & \frac{1}{N_C}\,
\sum_{N_c^{(\mbox{\tiny eik})}}
\left<0 \left| \Phi^{({\mathrm f}_c)}_{\xi_c}{}^\dagger(0)
\Phi_{\beta_c}^{({\mathrm f}_c)}{}^\dagger(0) \right| 
N_c^{(\mbox{\tiny eik})} \right>
\nonumber \\
& \ & \times\; \left< N_c^{(\mbox{\tiny eik})} \left|
\Phi^{({\mathrm f}_c)}_{\beta_c}(0)
\Phi_{\xi_c}^{({\mathrm f}_c)}(0) \right|  0 \right> \,
\delta\left(e_c -  e(N_c^{(\mbox{\tiny eik})}) \right)\, ,
\label{eikjetdef}
\eea
where the roles of the quark fields are taken by recoilless, lightlike 
Wilson lines.

We first observe that there is a certain ambiguity in the separation of jet and
soft functions.  
The ambiguity is exhibited clearly by 
a Laplace transform, where large $\nu$ is conjugate to small $e$.  
 Under the Laplace transform,
$d\sigma/de$ in Eq.~(\ref{firstfact})
factorizes into a simple product of jet and soft functions,
\begin{equation}
\tilde\sigma(Q,\nu) = \int_0\, de\, \e^{-\nu e}\, \frac{d\sigma(Q)}{de}
=\sigma_0(Q)\ 
 \tilde J_n(Q,\nu)\tilde J_{\bar n}(Q,\nu) \tilde S(Q/\nu)\, .
\label{factoredLaplace}
\end{equation}
Notice that dependence on the upper limit 
of the integral over $e$
is exponentially suppressed at large $\nu$.
The product on the right
in (\ref{factoredLaplace}) can be treated in different ways, depending on the task at
hand.  
The potential sources of double counting are eliminated in this case by
defining $\tilde S(Q/\nu)$ as the transform-space ratio
of the eikonal cross section to the product of eikonal jets,
\begin{equation}
\tilde S(Q/\nu) = \frac{\tilde\sigma^{\text{(eik)}}(Q,\nu)}{\tilde J^{\text{(eik)}}_n(Q,\nu)\tilde J^{\text{(eik)}}_{\bar n}(Q,\nu)}\, .
\label{eikonal}
\end{equation}
In Eq.~(\ref{eikonal}),
the eikonal cross-section and eikonal
 jet functions approximate well the wide-angle soft radiation  
in the full cross section and in the jet functions, respectively.
Thus the inverse eikonal jet functions in Eq.~(\ref{eikonal}) cancel the contributions of the soft radiation in the partonic jet functions in Eq.~(\ref{factoredLaplace}).
When expanded in the coupling, the denominators can be reinterpreted 
as sets of nested subtractions.

For the study of power corrections \cite{KS99}, it will be more useful to implement a slightly different organization, found by simply shifting the factors 
$\tilde J^{\text{(eik)}}_n(Q,\nu)\tilde J^{\text{(eik)}}_{\bar n}(Q,\nu)$
from the soft factors to the jets:
\begin{equation}
\tilde\sigma(Q,\nu) = \sigma_0(Q)\   \tilde{\cal J}_n(Q,\nu)\tilde{\cal J}_{\bar n}(Q,\nu) \tilde\sigma^{\text{eik}}(Q,\nu)\, .
\label{secondfactoredLaplace}
\end{equation}
where
\bea
\tilde{\cal J}_n(\nu) = \frac{\tilde J_n(Q,\nu)}{\tilde J^{\text{(eik)}}_n(Q,\nu)}\, , \qquad \tilde{\cal J}_{\bar n}(Q,\nu) = \frac{\tilde J_{\bar n}(Q,\nu)}{\tilde J^{\text{(eik)}}_{\bar n}(Q,\nu)}\, .
\label{subtractedjetfunctions}
\eea
  The  eikonal subtractions serve the same role as above in Eq.~(\ref{eikonal}),
 but now all double counting is subtracted from the jet functions. 
We will show in Sec.\ \ref{sec:double} that this
method of subtraction is
directly related to the ``zero-bin subtraction'' scheme \cite{0bin} in SCET. 
But first let us review the factorization of jet cross sections from the perspective of SCET.

\subsection{Factorization for event shapes in SCET}

The SCET analysis begins with
the expression for the distribution of event shape $e$ in QCD (the ``full theory"
from an effective theory point of view),
\begin{equation}
\frac{d\sigma}{de} = \frac{1}{2Q^2}\sum_N \abs{\bra{N} J^\mu(0)\ket{0}L_\mu}^2 
(2\pi)^4\delta^4(Q - p_N) \delta(e-e(N)),
\label{dsigmaQCD}
\end{equation}
where $J^\mu = \bar q\Gamma^\mu q$ is the production current in QCD.
The vector $L_\mu$ is the leptonic part of the amplitude for $\e^+ \e^-\rightarrow \bar q q$. SCET organizes QCD in an expansion in powers of a parameter $\lambda\sim\sqrt{\Lqcd/Q}$ \cite{SCET1,SCET2}. The modes in the effective theory we will use are collinear quarks and gluons, $\xi_{n,\bar n}$ and $A_{n,\bar n}$, and ultrasoft (usoft) gluons, $A_{us}$. These modes are distinguished by the scaling of the light-cone components 
$p=(n\cdot p,\bar n\cdot p,p_\perp)$ of their momenta, defined with respect to light-cone vectors $n,\bar n = (1,\vect{0_\perp},\pm 1)$. Ultrasoft momenta scale as $p_{us}\sim Q(\lambda^2,\lambda^2,\lambda^2)$. Collinear modes can have momenta with one of two possible scalings: $p_n\sim Q(\lambda^2,1,\lambda)$ or $p_n\sim Q(\lambda^4,1,\lambda^2)$, and similarly for $\bar n$-collinear momenta $p_{\bar n}$. The theory with collinear modes with the first scaling is called \SCETa, and with the second, \SCETb\ \cite{SCETII}. The typical transverse momenta of the collinear particles in the final state determines the correct choice of scalings. (For example, very narrow jets with $p_\perp\sim\Lqcd$ must be treated in \SCETb.) We begin by matching QCD onto \SCETa, and consider \SCETb\ when we wish to account for narrower jets.

At leading (zeroth) 
order in the  expansion in $\lambda$, the full theory current $J^\mu$ matches onto the \SCETa\ operators \cite{hardscattering},
\begin{equation}
j^\mu_{\omega,\omega'} = \bar\chi_{n,\omega} \Gamma^\mu \chi_{\bar n,\omega'},
\end{equation}
where $\omega,\omega'$ denote label momenta, and the jet fields $\chi_{n,\bar n}$ are given by
\bea
\chi_{n,\omega} = [W_n^\dag\xi_n]_{\omega}.
\label{chidef}
\eea
Here $W_n$ is a Wilson line of collinear gluons,
\begin{equation}
W_n(z) = P\exp\left[ig\int_{-\infty}^0 ds\,\bar n\cdot A_n(\bar n s+z)\right].
\label{collinearWilson}
\end{equation}
Label momenta are the ``large'' pieces of  collinear momentum. The momentum of a collinear mode splits into this label piece and a residual piece. More precisely, $p_n =  \tilde p_n + k$, where the label momentum $\tilde p_n = \bar n\cdot \tilde p_n\frac{n}{2} + \tilde p_n^\perp$ contains only the $\mathcal{O}(Q)$ piece of $\bar n\cdot p_n$ and the $\mathcal{O}(Q\lambda)$ piece of $p_n^\perp$ , and the residual momentum $k$ is of order $\Lqcd$ in all components. A collinear field with a label $\omega$ creates and destroys only modes with small $\mathcal{O}(\Lqcd)$ fluctuations about the momentum $\omega$.

The matching between the full and effective 
theories
is performed by matching matrix elements at a scale $\mu$:
\begin{equation}
\langle J^\mu\rangle_{\text{QCD}}(\mu) = C(\omega,\omega';\mu) \langle j^\mu_{\omega,\omega'}\rangle_{\text{SCET}}(\mu),
\end{equation}
where the labels $\omega,\omega'$ are summed over.
The coefficient 
$C(\omega,\omega';\mu)$ is the Wilson coefficient in this matching.    The combinations of fields and Wilson lines in these expressions
bear a close resemblance to the jet functions of Eq.~(\ref{jetdef}).

We can remove the coupling between ultrasoft gluons and the collinear fields in the \SCETa\ Lagrangian via the field redefinitions \cite{redef}, 
\begin{equation}
\xi_n = Y_n^\dag\xi'_n,\quad A_n= Y_n^\dag A'_n Y_n, \quad W_n= Y_n^\dag W'_n Y_n\, ,
\label{BPSredef}
\end{equation}
and similarly for the $\bar n$-collinear fields, using the ``outgoing'' \cite{Chay} Wilson line of ultrasoft gluons,
\begin{equation}
Y_n(z) = P\exp\left[ig\int_0^\infty ds\,n\cdot A_{us}( n s+z)\right].
\end{equation}
For instance, the term containing the  collinear quark-usoft gluon interaction becomes:
\begin{equation}
\bar\xi_n in\cdot D_{us} \xi_n \rightarrow \bar\xi'_n in\cdot\partial \xi'_n,
\end{equation}
where $iD_{us} = i\partial + gA_{us}$.
With this separation, we have established an effective theory that
captures the dynamics in the full theory for the two-jet event shape
distributions in the two-jet limit.  The identification of jet and soft
quanta precisely matches the ``leading regions" of the full theory,
as identified by analysis of momentum-space integrals for arbitrary
diagrams \cite{st78}.  
This analysis, of course, is
process-dependent, and the validity of this SCET holds
up to the same corrections that apply to Eq.~(\ref{firstfact}),
and is improvable by adding more jets, for example.

The redefined jet production current becomes
\begin{equation}
j^\mu_{\omega,\omega'} = \bar \chi'_{n,\omega} Y_n \Gamma^\mu Y_{\bar n}^\dag \chi'_{\bar n,\omega'},
\label{jetcurrentY}
\end{equation}
where the primes on the redefined fields $\chi'$ refer to the
use of $\xi'$ and $W'$ in Eq.~(\ref{chidef}).
This leads to a factorization of the differential cross-section in Eq.~(\ref{dsigmaQCD}),
analogous to Eq.~(\ref{firstfact}),
\begin{equation}
\frac{d\sigma}{de} = \abs{C(Q,-Q;\mu)}^2\ \int de_J\ \sigma_J(e_J;\mu) S(e-e_J;\mu),
\label{factorization}
\end{equation}
where the function $\sigma_J$ contains the collinear fields and states:
\begin{align}
\sigma_J(e_J;\mu) = \frac{1}{2Q^2}\frac{L^2}{3}\sum_{N_{J_n} N_{J_{\bar n}}}\abs{\bra{N_{J_n} N_{J_{\bar n}}} \bar\chi'_{n,Q} \Gamma^\mu \chi'_{\bar n,-Q}\ket{0}}^2(\mu) \delta(e_J - e(N_{J_n}N_{J_{\bar n}})),
\label{jetcrosssection}
\end{align}
where $L^2$ is the spin-averaged, squared leptonic amplitude, and
where the soft function $S$ contains the usoft fields:
\begin{equation}
S(e;\mu) = \frac{1}{N_C}\Tr \sum_{X_u}\abs{\bra{X_u}T[Y_n Y_{\bar n}^\dag]\ket{0}}^2(\mu)\delta(e-e(X_u)).
\label{shapefunction}
\end{equation}
We have split up the final state $N$ into the jets $N_{J_{n,\bar n}}$ and the usoft sector $X_u$, imagining them to live in clearly separated regions.
The notation in Eq.\ (\ref{factorization})
setting  the labels $\omega,\omega'$ of the collinear fields to $\pm Q$ means that the fields must create a quark and an antiquark jet with total label momenta $n\cdot\tilde p_{J_n} = \bar n\cdot\tilde p_{J_{\bar n}} = Q$, and $\tilde p_{J_n}^\perp=\tilde p_{J_{\bar n}}^\perp= 0$.

Before we proceed, let us consider briefly the scale dependence of the jet and soft functions. The matching from QCD to SCET is done at the scale $\mu=Q$, minimizing large logarithms in the Wilson coefficient $C(Q,-Q;\mu)$. The collinear matrix element is naturally evaluated at a scale $\mu_c\sim Q\lambda$, and the soft function  at $\mu_s\sim Q\lambda^2$, to minimize logarithms in those functions. The running between these scales and $\mu=Q$ is achieved by renormalization group evolution. The calculations required to do this in SCET are described in \cite{BS1,Trott}. In this paper, we focus not on the details of the perturbative matching and evolution, but rather on the properties of matrix elements at the scale $\mu\sim\Lqcd$ that must be treated as nonperturbative quantities.

For observables dominated by fairly wide jets with $p_\perp\sim Q\lambda\sim \sqrt{\Lqcd Q}$, the collinear jet function at $\mu_c=Q\lambda$ may be evaluated in perturbation theory, leaving only the soft function at $\mu_s=\Lqcd$ as a nonperturbative function. For observables in which narrow jets with $p_\perp\sim\Lqcd$ contribute heavily, 
the collinear matrix elements are naturally evaluated at the scale $\mu_c = \Lqcd$. Between the scales $Q\lambda$ and $Q\lambda^2$, the theory \SCETa\ at the higher scale must be matched onto \SCETb\ at the lower scale. The form of the collinear operator (\ref{jetcurrentY}) remains the same; only the momentum scaling of the collinear fields changes. A collinear matrix element in \SCETb\ is a nonperturbative quantity. In Sec.~\ref{sec:power}, we compare the relative sizes of nonperturbative corrections to event shape distributions from the soft and
 jet functions. For the observables of most interest to us, the power corrections from the soft function will dominate. 

Returning now to the jet function in Eq.~(\ref{jetcrosssection}), we can further separate the matrix elements involving the jet
quanta into the form:
\bea
\label{sepjets}
\sigma_J(e_J;\mu) &=& \frac{1}{2Q^2}\frac{L^2}{3} 
 {\rm Tr}\; \big [\Gamma_\mu^\dagger 
\sum_{N_{J_n} } \bra{0 } \chi'_{n,Q} \ket{N_{J_n}} \bra{N_{J_n }} \bar\chi'_{n,Q} \ket{0} \\
&\ & \hspace{10mm} \times  \Gamma^\mu
\sum_{N_{J_{\bar n}}} \bra{0} \bar \chi'_{\bar n,-Q}\ket{N_{J_{\bar n}}}\, \bra{N_{J_{\bar n}}}  \chi'_{\bar n,-Q}\ket{0}
\big](\mu) \delta(e_J - e(J_n)-e(J_{\bar n}) )\, .\nonumber
\eea
The individual squares of matrix elements are closely related to 
the jet functions in factorization-based treatments of semi-inclusive
cross sections, Eq.~(\ref{jetdef}) \cite{BKS03}.   In particular, the operator content,
including the presence of the Wilson lines $W_n$, is essentially equivalent.
The choice of vector ($\xi_c$ in (\ref{jetdef})) is somewhat different. 
Another difference is  that here the collinear fields are fully
separated from usoft partons, so that the sums over explicit jet
final states $\ket{N_{J_{n,\bar n}}}$ contain no usoft lines at all, while
correspondingly, the virtual states that enter the jet matrix elements
have no soft lines.   In the soft function as well, jet lines appear
neither in the final state $\ket{X_u}$ nor in the matrix elements.

In SCET, this bookkeeping is built in
 by a ``zero-bin subtraction" prescription for collinear fields \cite{0bin}.  
 For each collinear field in the SCET Lagrangian, there is a sum over all label momenta, which must always be nonzero, to avoid overlap with usoft modes. Thus, regions of collinear loop diagrams or phase space integrals where a collinear particle can become ultrasoft must always be subtracted off. 

Each particle in the final state may be placed in a bin determined by its label momentum, leaving it with a residual momentum inside the bin, while those with zero label fall into the zero bin, and are assigned to the usoft sector $X_u$. The couplings of usoft particles to the states $N_{J_{n,\bar n}}$ have already been factored out and accounted for in the soft function $S$.

A natural question at this point is whether the jet and soft functions so defined
are individually infrared safe.  The answer is yes, because infrared
safety does not require Lorentz invariance, only the hermiticity of
the interaction Hamiltonian, and a sufficiently smooth shape
function \cite{st79}.  In our case, the cancellation of soft and
collinear singularities in leptonic annihilation processes can be carried
out for fixed values of one of the components of light-cone momentum \cite{st78}.
We thus expect infrared safety for each of the SCET jet and soft functions.

\section{Eliminating Double Counting}
\label{sec:double}

In this section, we continue our comparison of the
SCET treatment of event shapes to analyses based
directly on factorization \cite{KS99,BKS03}, concentrating on the
elimination of double counting.
The observations below 
are relatively simple, but to our knowledge they have not
yet been made in the literature in this context.

Proofs of factorization in perturbative QCD typically make use of 
Ward identities and subtractions to avoid double counting 
of leading configurations while maintaining infrared safety \cite{pqcdfact,BKS03}. 
Unlike their SCET analogs, the matrix elements in Sec.\ \ref{3a} above
have no restrictions on their momentum
integrals.  The jet functions thus include a considerable contribution
from what in SCET are classified as usoft gluons.\footnote{They also include quanta collinear to the opposite-moving
jet.  In SCET, these quanta are simply not present in the jet lines.
In the factorized jet functions of Eq.~(\ref{jetdef}), although present
they do not produce collinear singularities because the
Wilson lines $\Phi_{\xi_c}$
are defined with respect to vectors $\xi_c$ that are not lightlike.}
 The soft function $S$ was correspondingly 
  defined to avoid double-counting of collinear gluons by the
  soft function, and wide-angle soft gluons by the jet functions.
As noted above, the soft function is
constructed from Wilson lines in a manner precisely 
analogous to the SCET soft function, Eq.~(\ref{shapefunction}).  
From Eqs.~(\ref{secondfactoredLaplace})
and (\ref{subtractedjetfunctions}), 
with double-counted soft modes subtracted from the collinear jets, we see that the
eikonal cross section 
 of (\ref{eikdef})
is exactly the same as the SCET soft function (\ref{shapefunction}), except that
the former does not include an explicit limitation to usoft final states, $X_u$.
 
The subtraction of double-counted soft modes from the jet functions is precisely what is achieved in SCET by the zero-bin subtraction for collinear fields (cf. Eqs.~(74) and (75) in Ref. \cite{0bin}).
 Although there is a restriction to ultrasoft states in (\ref{shapefunction}), the phase space integrals implicit in the sum still cover the entire region of momentum space, so that the eikonal soft function in QCD and the SCET shape function (\ref{shapefunction}) are practically equivalent. It is not inconsistent to integrate up to infinite momenta in a usoft integral. One is first taking the limit $\lambda\rightarrow 0$, {\it i.e.} $Q\rightarrow\infty$, then integrating up to infinity the usoft momenta, which remain always formally much smaller than $Q$.\footnote{We thank A. Manohar for discussions on this point. See related  remarks in Ref. \cite{BLMW}
 .} It is important only that the soft functions for the observables we calculate are dominated by those particles which have ``truly'' usoft momenta. Below, we will examine the validity of this assumption for the event shapes we consider. First, let us elaborate on the connection between double-counting subtraction procedures in QCD and SCET.

\subsection{Zero-bin subtractions in factorization}

The leading-order (in $\lambda$) Lagrangian of SCET contains several parts (see, \emph{e.g.}, Ref.~\cite{redef}). First, the purely ultrasoft Lagrangian for quarks and gluons is identical to that in QCD:
\begin{equation}
\mathcal{L}_{\text{us}} = \bar q_{\text{us}} i{\Dslash}_{\text{us}} q_{\text{us}} - \frac{1}{2}\Tr G^{\mu\nu}G_{\mu\nu} + \mathcal{L}_{\text{us}}^{\text{g.f.}},
\label{usoftLagrangian}
\end{equation}
where the ultrasoft field strength,
 $G^{\mu\nu} = \frac{i}{g}[D_{\text{us}}^\mu,D_{\text{us}}^\nu]$, 
is given in terms of the ultrasoft covariant derivative,
 $iD_{\text{us}}^\mu = i\partial^\mu + gA^\mu_{\text{us}}$.  
 The usoft gauge fixing and ghost terms
 are represented by $\mathcal{L}_{\text{us}}^{\text{g.f.}}$.
  Meanwhile, the Lagrangian for collinear quarks is
\bwt
\begin{equation}
\mathcal{L}_c^{\text{(q)}}  = \bar\xi_{n,p'}\left[in\cdot D_{\text{us}} + gn\cdot A_{n,q} + (\Pslash_\perp + g\Aslash^\perp_{n,q}) 
W\frac{1}{\mathcal{\bar P}}W^\dag(\Pslash_\perp + g\Aslash^\perp_{n,q'})\right]\frac{\bnslash}{2}\xi_{n,p},
\end{equation}
\ewt
where $\mathcal{\bar P}$ and $\mathcal{P}_\perp$ are label momentum operators 
which pick out the $\mathcal{O}(Q)$  piece of the $\bar n\cdot p$ component and the $\mathcal{O}(Q\lambda)$ piece of the $p_\perp$ component of the collinear momenta.
Here $W$ is the Wilson  line defined
in Eq.~(\ref{collinearWilson}), organizing collinear gluons $\bar{n}\cdot A$,
each of which are leading power in $\lambda$ \cite{labelops}.
There is also an implicit sum over all labels and conservation of label momentum in each term 
\cite{labelops}. Finally, the collinear gluon Lagrangian is
\begin{equation}
\mathcal{L}_c^{\text{(g)}} = \frac{1}{2g^2} \Tr\left\{[i\mathcal{D}^\mu + g A_{n,q}^\mu, i\mathcal{D}^\nu+ gA_{n,q'}^\nu]^2\right\} + \mathcal{L}_c^{\text{g.f.}},
\end{equation}
where $i\mathcal{D}^\mu = \frac{n^\mu}{2}\mathcal{\bar P} + \mathcal{P}_\perp^\mu + \frac{\bar n^\mu}{2}in\cdot D_{\text{us}}$, and $\mathcal{L}_c^{\text{g.f.}}$ contains the collinear gauge fixing 
and ghost terms.

The component $n\cdot A_{\text{us}}$ of the usoft gluon field appears in the collinear quark and gluon Lagrangians above. The field redefinition (\ref{BPSredef}) removes this interaction, through the identities $in\cdot D_{\text{us}}Y_n^\dag = 0$ and $i\mathcal{D}^\mu = Y_n^\dag i\mathcal{D}_{(0)}^\mu Y_n$,
where $i\mathcal{D}_{(0)}^\mu = \frac{n^\mu}{2}\mathcal{\bar P} + \mathcal{P}_\perp^\mu + \frac{\bar n^\mu}{2}in\cdot \partial$.  This gives
the redefined collinear quark Lagrangian,
\bwt
\begin{equation}
{\mathcal{L}_c'}^{\text{(q)}} = \bar\xi'_{n,p'}\left[in\cdot \partial + gn\cdot A'_{n,q} + (\Pslash_\perp + g{\Aslash'}^\perp_{n,q}) W\frac{1}{\mathcal{\bar P}}W^\dag(\Pslash_\perp + g{\Aslash'}^\perp_{n,q'})\right]\frac{\bnslash}{2}\xi'_{n,p},
\label{calLprimedef}
\end{equation}
\ewt
and collinear gluon Lagrangian, 
\begin{equation}
{\mathcal{L}_c'}^{\text{(g)}} = \frac{1}{2g^2} \Tr\left\{[i\mathcal{D}_{(0)}^\mu + g {A'}_{n,q}^\mu, i\mathcal{D}_{(0)}^\nu+ g{A'}_{n,q'}^\nu]^2\right\} + {\mathcal{L}_c'}^{\text{g.f.}}\, .
\end{equation}

It is in the theory described 
by the redefined Lagrangian $\mathcal{L}_{\text{(c)}}'$ that we derived
factorization for the two-jet event shape distributions. As given in Eq.~(\ref{factorization}), the full distributions are convolutions of soft and 
separate jet functions for the two jets, which are
easily abstracted from Eq.~(\ref{sepjets}). For example, in terms of the collinear SCET
quark field, $\xi_n'$ in (\ref{calLprimedef}), we can define a 
jet function $J^c_n(e)$, as
\begin{equation}
J^c_n(e)\left(\frac{\nslash}{2}\right) = \sum_{N_{J_n}}\abs{\bra{N_{J_n}}[\bar\xi'_n {W_n'}]_{p_n}\ket{0}}^2_{\mathcal{L}_c'}\delta(e - e(N_{J_n})),
\label{Jcdef}
\end{equation}
where the total label momentum of the collinear fields must equal the label momentum $\tilde p_n$ of the states $N_{J_n}$, which are free of quanta in the usoft sector. 
The right-hand side is
proportional to the matrix $\nslash/2$, as shown, and 
$J_n^c(e)$ is a remaining scalar function.  Here we include a subscript on the
matrix elements to emphasize the Lagrangian in question.
This jet function contains contributions only 
from ``truly'' collinear particles with nonzero label momenta, and all usoft particle contributions are accounted for in the soft function $S$. 
Restricting collinear label momenta to nonzero values avoids double-counting of usoft contributions in these two functions. It may be
convenient, however, to allow integrals over collinear momenta to include the ``zero-bin''. One must then define a prescription to subtract it out again. This can
 be achieved by the zero-bin subtraction procedures illustrated by Manohar and Stewart in Ref.~\cite{0bin}, or by the method of factoring out ``eikonal'' jet functions described by  Ref.~\cite{BKS03}. 
 
 In order to establish the connection between the 
 zero-bin and eikonal jet methods for subtraction, let us take another look 
at the collinear quark Lagrangian in SCET, now with sums over labels explicit:
\bwt
\begin{equation}
{\mathcal{L}_c'}^{(q)}(x) = \sum_{\tilde p,\tilde p'\not = 0} \e^{i(\tilde p'-\tilde p)\cdot x}\bar \xi'_{n,p'}(x)\Bigl[in\cdot\partial + g\sum_{\tilde q\not =0}\e^{-i\tilde q\cdot x}n\cdot A'_{n,q}(x)\Bigr]\frac{\bnslash}{2}\xi'_{n,p}(x)\, .
\end{equation}
\ewt
We have dropped for now the terms with $\Aslash^\perp$ gluons, since they are irrelevant for the remainder of this discussion\footnote{The perp gluons in the zero-bin overlap with usoft gluons, but these interact with collinear fields only at subleading order in $\lambda$.}. The tildes denote label momenta containing only the large $\mathcal{O}(Q)$ and $\mathcal{O}(Q\lambda)$ pieces, 
$\tilde p^\mu = \bar{n}\cdot \tilde p\frac{n^\mu}{2} + \tilde p_\perp^\mu$, and the remaining $x$ dependence of collinear fields fluctuates on a scale described by residual momenta $k$ of order $Q\lambda^2$.  The sums over collinear labels are restricted to nonzero values to avoid double counting the usoft modes, and we have shown explicitly the phases 
that enforce label momentum conservation. 
As in the zero-bin prescription of Ref.~\cite{0bin},
 the sums over labels can be allowed to include the zero bin if we add a term that subtracts it out again. Let us worry about this only for the gluon field, since the interaction of usoft quarks with collinear fields is subleading in $\lambda$.
These steps can be achieved 
beginning with a Lagrangian $\mathcal{L}''$ that includes zero-bin collinear gluons:
\bwt
\begin{equation}
{\mathcal{L}_c''}^{(q)}(x) = \sum_{\tilde p,\tilde p'\not = 0} \e^{i(\tilde p'-\tilde p)\cdot x}\bar \xi''_{n,p'}(x)\Bigl[in\cdot\partial + g\sum_{\tilde q}\e^{-i\tilde q\cdot x}n\cdot A''_{n,q}(x)\Bigr]\frac{\bnslash}{2}\xi''_{n,p}(x)\, .
\end{equation}
\ewt
We then perform a field redefinition on (nonzero-bin) collinear fields:
\begin{equation}
\xi''_{n,p} = U_n^\dag \xi'_{n,p}\, , \qquad A''_{n,p} = U_n^\dag A'_{n,p} U_n\, , \qquad
W''_n=  U_n^\dag \tilde{W}'_n U_n\, , \qquad
\label{0binredef}
\end{equation}
where $U_{n}$ is a ``zero-bin'' Wilson line:
\begin{equation}
U_n(x) = P\exp\left[ig\int_0^\infty ds\,n\cdot A''_{n,0}(ns+x)\right],
\end{equation}
mimicking the field redefinition (\ref{BPSredef}) with the usoft Wilson line.
The Lagrangian obtained from ${\mathcal{L}''_c}^{(q)}$ by the redefinition (\ref{0binredef}) is precisely ${\mathcal{L}_c'}^{(q)}$, with the $\tilde q=0$ term subtracted off. The redefinition of the gluon field removes the zero-bin gluons from the corresponding collinear gluon Lagrangian.
Note, however, that it does not yet remove zero-bin 
$\bar{n}\cdot A_n$ fields
in $W''$.  Thus, the Wilson line $\tilde {W}'$ is not yet $W'$, when the latter
is defined to be free of zero-bin gluons.

The jet function $J^c_n(e)$ of Eq.~(\ref{Jcdef})
is, as indicated, calculated in the theory described by $\mathcal{L}_c'$ in order to avoid double-counting the usoft modes. However, if we for convenience decide to include the zero-bin in all the collinear momentum integrals, we are actually calculating
a slightly different jet function,
 $J_n$, in the theory described by $\mathcal{L}_c''$. 
  Via the field redefinition (\ref{0binredef}) we can move back to the Lagrangian $\mathcal{L}_c'$, but then we must include the $U_n$ Wilson line in the current:
\begin{align}
J_n(e) \left(\frac{\nslash}{2}\right) &= \sum_{N_{J_n}}\abs{\bra{N_{J_n}}[\bar \xi''_n{W_n''}]_{p_n}\ket{0}}_{\mathcal{L}_c''}^2\delta(e-e(N_{J_n})) \nonumber \\
&=  \sum_{N_{J_n}}
\abs{\bra{N_{J_n}}[\bar \xi'_n {\tilde{W}_n'}]_{p_n}U_n\ket{0}}_{\mathcal{L}_c'}^2\delta(e-e(N_{J_n})).
\end{align}
In the Lagrangian $\mathcal{L}_c'$, however, the zero-bin gluons do not interact with the collinear fields at all. 
Once the $\bar{n}\cdot A_{n,0}$ zero-bin gluons have been removed from $\tilde{W}'$,
 this jet function factorizes into ``purely'' collinear and ``zero-bin'' parts.  
To accomplish this,
let us  also split the collinear Wilson line $\tilde{W}'_n$ into its zero-bin and purely collinear parts, 
\bea
\tilde{W}'_n(z) =  W_n'(z)\, \Omega_n(z)\, ,
\eea
where
\begin{equation}
\Omega_n(z) = P\exp\left[ig\int_{-\infty}^0 ds\,\bar n\cdot A_{n,0}(\bar n s+z)\right],
\end{equation}
and $W_n'$ is the collinear Wilson line with restrictions to nonzero label momenta on all gluons.
Then the alternative (${\cal L}''$) 
jet function factorizes into:
\bwt
\begin{align}
J_n(e)\left(\frac{\nslash}{2}\right) &= \sum_{N_{J_n^c},N^{\text{(eik)}}}\abs{\bra{N_{J_n^c}}[\bar\xi'_n {W'_n}]_{p_n}\ket{0} \langle N^{\text{(eik)}}\rvert \Omega_n U_n\ket{0}}^2_{\mathcal{L}_c'}
\delta(e - e(N_{J_n^c}) - e(N^{\text{(eik)}})) \nonumber \\
&= \int de_{c} J_n^c(e_{c}) J_n^{\text{(eik)}}(e - e_c)\left(\frac{\nslash}{2}\right),
\end{align}
\ewt
where $J_n^c$ is the jet function calculated in the purely-collinear theory described by $\mathcal{L}_c'$, and $J_n^{\text{(eik)}}$ is the eikonal jet function. The collinear fields in $J_n^c$ can only produce gluons with nonzero label momenta, while the Wilson lines $\Omega_n, U_n$ in $J_n^{\text{(eik)}}$ produce gluons in the zero bin. It is closely related to the eikonal jet function defined
above,  in Eq.~(\ref{eikjetdef}).  

By taking the Laplace transform of the jet functions in this convolution, we may solve for the purely collinear jet function:
\begin{equation}
\tilde J_n^c(\nu) = \frac{\tilde J_n(\nu)}{\tilde J_n^{\text{(eik)}}(\nu)},
\end{equation}
which is precisely Eq.~(\ref{subtractedjetfunctions}). This relation tells us that we may calculate the jet function on the left, $J_n^c$, by calculating instead the jet function on the right, $J_n$, which include the zero-bins in collinear integrals, if we then subtract out the double-counted usoft contributions with the eikonal jet function, which is precisely the prescription
of Ref.~\cite{BKS03}. 
We emphasize that the alternative SCET jet function, $J_n$, is directly analogous to the
 pQCD jet function defined in Eq.~(\ref{jetdef}).   There remain technical differences associated
 with the states $\ket{N_{J_c}}$ in (\ref{jetdef}), which can also contain energetic lines collinear
to the opposite-moving jet, for example.  
As noted above,
the effect of such lines is perturbatively calculable,
so that these remaining differences are perturbatively calculable.
We have thus shown its equivalence to the 
 zero-bin subtraction procedure of Ref.~\cite{0bin} in SCET.

\subsection{The ultrasoft region in SCET}
\label{usoftregion}

For the purposes of our discussion below, we need to consider in
some detail the precise definition of our usoft region. In each of the
two-jet event shapes of Sec.\ 2, the contributions of particles in
the far forward and backward regions are suppressed exponentially
in the absolute value of
their rapidity, and hence as a power of their larger light-cone momentum (order $Q$).
The contributions of jet-like particles at low (order $\Lambda$) transverse
momenta are suppressed by a power of $Q$ compared to those emitted
with similar low transverse momenta at wide angles.   For this reason,
to leading powers in $1/(eQ)$, we need not include separate nonperturbative
functions for the jets \cite{KS95}, which 
may be expected to
enter beginning at the
level of $1/(eQ^{1+b})$, where $b$ depends on the
event shape at hand.  For the angularities, $b=1-a$.  
We will give a more formal argument for this result in the next section.
The soft shape function organizes all powers
like $1/(eQ)^n$, while neglecting all further suppression by powers of $Q$.

Our analysis of power corrections from the soft function will depend crucially on the boost invariance of the usoft Wilson lines appearing in $S$, and, in the sum over usoft states in (\ref{shapefunction}), we integrate over all momenta. 
With this in mind, we will want our ultrasoft region formally to include
all gluons with small transverse momenta but boosted to arbitrarily large rapidities. The light-cone components $k^\pm$ of their momenta are then, strictly speaking, larger than the typical usoft scaling $Q\lambda^2$. Formally, the usoft modes in SCET do cover all momenta up to the scale $\mu=Q$, although the effective theory Lagrangian is a good approximation to the dynamics only of the ``truly'' usoft particles. As long as we pick observables to which only these truly usoft particles contribute significantly, it is safe to include the highly-boosted particles as well in the usoft sector of the theory. Indeed, for event shapes such as the angularities, the contribution of particles in the far-forward and far-backward regions are power-suppressed. 

Consider for example the shape function (\ref{shapefunction}) for the angularities as defined in Eq.~(\ref{shapedefs}).  Do the usoft modes in SCET correctly describe the dynamics of all the small transverse-momentum particles that make a non-negligible contribution to the observable $\tau_a$? Let us say that if the exponential factor, $\exp[-\abs{\eta_i}(1-a)]$, for particle $i$ is of order $\lambda$ or smaller, then its contribution is negligible. Then the largest-rapidity particle making a non-negligible contribution to the event shape has:
\begin{equation}
\abs{\eta_i}\sim -\frac{1}{1-a}\ln\lambda,
\end{equation}
or, defining the rapidity as 
$\eta = \frac{1}{2}\ln(\bar n \cdot k / n \cdot k)$,
\begin{equation}
\max\biggl(\frac{\bar{n}\cdot k}{n\cdot k},\frac{n\cdot k}{\bar{n}\cdot k}\biggr)\sim \lambda^{-\frac{2}{1-a}}.
\label{maxeq}
\end{equation}
Now, because $n\cdot k\, \bar{n}\cdot k \sim k_\perp^2$, and for
usoft particles, $k_\perp\sim Q\lambda^2$,  
Eq.\ (\ref{maxeq}) implies that
usoft particles with the larger of their light-cone momenta up to the order $Q\lambda^{2-\frac{1}{1-a}}$ contribute non-negligibly to the event shape $\tau_a$. For $a<1/2$, this light-cone momentum is still smaller than the corresponding component of a collinear momentum,
 which is order $Q$ ({\it e.g.} for the thrust, $a=0$.
  With $Q\sim 100\text{ GeV}$ and $\Lqcd\sim 1\text{ GeV}$, we have $\lambda = 0.1$, so the largest usoft light-cone momentum that contributes to $\tau_0$ is $10\text{ GeV}$, still well below $Q$.) As long as this hierarchy of scales holds for the large light-cone components of usoft and collinear momenta,  the only component of usoft gluons that interacts with collinear modes in the $n$ direction in the leading-order SCET Lagrangian is the $n\cdot A_{us}$ component ($\bar n\cdot A_{us}$ in the $\bar n$ direction), so that these interactions can be removed by the field redefinitions with the Wilson lines $Y_{n,\bar n}$. This guarantees the form of our usoft shape function (\ref{shapefunction}), and ensures that only the ``truly'' usoft particles contribute to the sum. We can extend the range of allowed values of $a$ beyond $a<1/2$ (but only up to $a<1$) by relaxing our criterion for the size of ``non-negligible" terms ({\it i.e.}\ allowing $\exp[-\abs{\eta_i}(1-a)]$ to be larger than $\lambda$ but smaller than 1).
For the $C$-parameter, we need $1/\cosh\eta_i < \lambda$, which for the values chosen above translates to 
$n\cdot k,\bar n\cdot k\lesssim 20\text{ GeV}$. 

Thus, in the following we may safely incorporate the power-suppressed
 contributions
of the very far-forward and far-backward radiation in the
SCET shape function, Eq.~(\ref{shapefunction}), and identify it with the eikonal cross section, Eq.~(\ref{eikdef}),
evaluated at the corresponding scale.
This means that in both SCET and full QCD 
we may treat the sum over states in the
shape function as boost-invariant.  It is this result that will lead us to
demonstrate  universality properties below.

\section{Identifying Power Corrections }
\label{sec:power}

We are now ready to identify the power corrections that arise naturally 
when the soft and jet functions  are evaluated at scales of order $\Lqcd$.
In doing so, we set aside issues of perturbative resummation and
of matching, treated in full QCD for the angularities in Ref.~\cite{BKS03},
and very recently in SCET for 
the closely-related jet cross sections by Refs.~\cite{BS1,Trott}.
In the discussion of this section, we will find useful a variant of the 
energy flow operators introduced in in Refs.\ \cite{flowops}
and applied in this context by Ref.~\cite{KS99}.  This operator
is clearly also closely related to the energy-energy correlations
of Refs.\ \cite{BBEL}.

As mentioned earlier, the typical transverse momenta of the collinear particles in the jets which contribute to a given observable determine whether they should be treated in the theory \SCETa, in which collinear momenta scale as 
$p_c = (n\cdot p_c,\bar n\cdot p_c,p_c^\perp)\sim Q(\lambda^2,1,\lambda)$ or $Q(1,\lambda^2,\lambda)$, recalling that $\lambda\sim\sqrt{\Lqcd/Q}$. The typical virtuality of such particles being $p_c^2 = Q\Lqcd$, jet functions in this theory can be calculated perturbatively. However, some event shapes may weight much narrower jets more heavily, in which jet constituents with transverse momenta of order $Q\lambda^2\sim\Lqcd$ become important. These degrees of freedom must be treated as collinear particles in \SCETb, in which collinear momenta scale as $p_c\sim Q(\lambda^4,1,\lambda^2)$ or $Q(1,\lambda^4,\lambda^2)$. These particles have virtualities of order $p_c^2\sim\Lqcd^2$, and so give rise to nonperturbative effects in addition to those from soft particles. For such event shapes, nonperturbative power corrections to the jet functions may compete
with (or even dominate) those in the soft functions.

We will now show how these results can be justified for the
event shapes in question.  We will  give our arguments  in terms of SCET matrix elements,
keeping in mind that they can be 
presented in terms of
 matrix elements in full QCD in a similar manner.
For the $C$-parameter and angularities $\tau_a$ with $a<1$, the dominant power corrections (of the order $\Lqcd/Q$) will come only from the effect of usoft particles whose momenta are of $\mathcal{O}(Q\lambda^2)$. Power corrections from collinear particles will be found to scale as $(\Lqcd/Q)^{2-a}$, which then dominate for $a\geq 1$. However, for $a\geq 1$, there are also other power corrections, for example, due to the shift in the thrust axis itself caused by the soft radiation \cite{broadening1,broadening2,BKS03}. The inclusion of these effects, while necessary for a complete treatment of power corrections to $\tau_a$ with $a \geq 1$, is outside the scope of this paper.

Consider the distribution of an event shape of the form in Eq.~(\ref{eventshape}), given in SCET by Eqs.~(\ref{factorization}--\ref{shapefunction}). 
The collinear cross-section (\ref{jetcrosssection}) is, writing out the general
event shape of Eq.~(\ref{eventshape}) in the delta function explicitly,
\begin{equation}
\sigma_J(e_J;\mu_c) = \frac{1}{2Q^2}\sum_{N_{J_n}N_{J_{\bar n}}}\abs{\bra{N_{J_n}N_{J_{\bar n}}}
\bar\chi_{n,Q}\Gamma^\mu\chi_{\bar n,-Q}\ket{0}}^2(\mu_c)\delta\Bigl(e_J - \frac{1}{Q}\sum_{i\in N_{J_n}N_{J_{\bar n}}}\abs{\vect{p}_i^\perp}f_e(\eta_i)\Bigr),
\label{collineardelta}
\end{equation}
while the soft function is
\begin{equation}
S(e;\mu_s) = \frac{1}{N_c}\Tr\sum_{X_u} \abs{\bra{X_u}Y_n \overline Y_{\bar n}\ket{0}}^2(\mu_s)\delta\Bigl(e - \frac{1}{Q}\sum_{i\in X_u}\abs{\vect{k}_i^\perp}f_e(\eta_i)\Bigr),
\label{softdelta}
\end{equation}
where we have now chosen to denote explicitly the dependence of the jet and soft functions on the scales $\mu_c,\mu_s$. 
Also, we have suppressed the factor associated with the leptonic part, and we have 
removed the time-ordering operator that was in the soft function in Eq.~(\ref{shapefunction}) by using the Wilson line $\overline Y_{\bar n}$, where the bar denotes the anti-fundamental representation of $SU(N_c)$ \cite{BLMW}.
For event shapes such as $\tau_a$ for $a<1$, the collinear scale $\mu_c$ can be chosen at a perturbative scale $\mu_c\sim Q\lambda$, and we are in \SCETa. For $a>1$, the event shapes pick out narrower jets so that the collinear scale is determined to be of order $\mu_c\sim\Lqcd$, putting us in \SCETb, where the jet function is nonperturbative.

We may express the delta functions in Eqs.~(\ref{collineardelta}) and (\ref{softdelta}) in operator form by making use of a transverse energy flow operator, defined by its action on states $N$:
\begin{equation}
\label{TEF}
\mathcal{E}_T(\eta)\ket{N(k_i)} = \sum_{i\in N}\abs{\vect{k}_i^\perp}\delta(\eta - \eta_i)\ket{N(k_i)},
\end{equation}
where the sum is over the particles $i$ in state $N$. This is equivalent 
to the energy flow operators discussed in Refs.~\cite{KS99,flowops,flow}. 
In terms of this operator, the collinear and soft functions (\ref{collineardelta},\ref{softdelta}) 
can be written as
\begin{align}
\sigma_J(e_J;\mu_c) &= \frac{1}{2Q^2}\sum_{N_{J_n}N_{J_{\bar n}}}\bra{0}
\bar\chi_{\bar n,-Q}\bar\Gamma^\mu\chi_{n,Q}\delta\Bigl(e_J - \frac{1}{Q}\int_{-\infty}^\infty d\eta\,f_e(\eta)\mathcal{E}_T(\eta)\Bigr) \ket{N_{J_n}N_{J_{\bar n}}} \nonumber \\
&\qquad\qquad\qquad\times\bra{N_{J_n} N_{J_{\bar n}}}
\bar\chi_{n,Q}\Gamma^\mu\chi_{\bar n,-Q}\ket{0},
\end{align}
and
\begin{equation}
S_{e}(e;\mu_s) = \frac{1}{N_C}\Tr\sum_{X_u}\bra 0 \overline{Y}_{\bar n}^\dag Y_n^\dag\delta\left(e - \frac{1}{Q}\int_{-\infty}^\infty d\eta\,
f_e(\eta)\mathcal{E}_T(\eta)\right) \ket{X_u}\bra{X_u}Y_n\overline{Y}_{\bar n}\ket{0}\, .
\label{shapewithXu}
\end{equation}
We can expand the delta functions in power series to identify the power corrections. 
If we first factor out the overall, canonical factor of $1/e$, 
shared with perturbation theory, and assume that the matrix elements are
of the order
of the momentum components of the usoft gluons,
$Q\lambda^2  \sim \Lambda_{\rm QCD}$,
we  derive a power series in $\Lambda_{\rm QCD}/(eQ)$.  
Indeed, the purpose of event shape functions is to 
organize all terms in this series when $\Lambda_{\rm QCD}/(eQ)\sim 1$
and all such power corrections are comparable.  These power corrections are particularly clearly
exhibited by Laplace transforms, Eq.~(\ref{factoredLaplace}), of the soft function
at low scales \cite{KS99},
\bea
\tilde S_{e}(\nu;\mu_s)
&=& \int_0\ de \exp[-\nu\, e]\ S_{e}(e;\mu_s) 
\nonumber\\
&=&
\frac{1}{N_C}\Tr \bra 0 \overline{Y}_{\bar n}^\dag Y_n^\dag
\exp \left[  - \frac{\nu}{Q}\int_{-\infty}^\infty d\eta\,f_e(\eta) \mathcal{E}_T(\eta)\right ] 
Y_n\overline{Y}_{\bar n}\ket{0}\, ,
\label{shapewithoutXu}
\eea
where we have summed over the complete set of intermediate states
in (\ref{shapewithXu}), as argued in Sec.~\ref{usoftregion}.   Expanding the exponential, we find 
a series in powers of the Laplace variable $\nu$ divided by $Q$.

For the soft function, according the discussion in the previous section, the sum over usoft states is unrestricted, as is the integral over rapidities inside the delta function. In the collinear function, choosing $e_J$ to be close to the two-jet limit $e_J=0$, or specifying a jet definition to pick out two-jet events, restricts the phase space integrals in the collinear cross-section to those with large rapidities, effectively limiting the range of the rapidity integral as well. The rapidities $\eta$ can be written in terms of the light-cone momenta of final state partons,
$\eta = \frac{1}{2}\ln(\bar n\cdot p/n\cdot p)$.
For a usoft parton, the ratio 
$n\cdot p_{us}/\bar{n}\cdot p_{us} \sim 1$, as all momentum components are $\mathcal{O}(Q\lambda^2)$, so $\abs{\eta} \sim 0$, while for collinear partons in \SCETb, one light-cone component is $\mathcal{O}(Q)$ while the other is $\mathcal{O}(Q\lambda^4)$. Thus, 
$n\cdot p_c/\bar{n}\cdot p_c \sim \lambda^4$ or $\lambda^{-4}$, so $\eta \sim \pm\ln\lambda^2$. Consider what this implies for the collinear and soft functions 
in the case of the angularities. The function $f_{e}(e)$ for $e=\tau_a$ is $f_{\tau_a}(\tau_a) = \e^{-\abs{\eta}(1-a)}$. In the usoft function, this factor is of $\mathcal{O}(1)$. In the collinear integral, the phase space restrictions limit the rapidity integral to $\abs{\eta}\gtrsim\ln(1/\lambda^2)$, so that the collinear function is effectively
\begin{align}
\sigma_J(\tau_a) &= \frac{1}{2Q^2}\sum_{N_{J_n}N_{J_{\bar n}}}\bra{0}
\bar\chi_{\bar n,-Q}\bar\Gamma^\mu\chi_{n,Q}\delta\Bigl(\tau_a - \frac{2}{Q}\int_{\ln\frac{1}{\lambda^2}}^\infty d\eta\,\e^{-(1-a)\eta}\mathcal{E}_T(\eta)\Bigr) \ket{N_{J_n}N_{J_{\bar n}}} \nonumber \\
&\qquad\qquad\qquad\times\bra{N_{J_n} N_{J_{\bar n}}}
\bar\chi_{n,Q}\Gamma^\mu\chi_{\bar n,-Q}\ket{0}.
\end{align}
Although we cannot compute these nonperturbative matrix elements at the scale
$\mu_c\sim Q\lambda^4$, we can estimate their dependence on $\lambda$ from dimensional analysis.
Matrix elements of powers of the operator $\mathcal{E}_T(\eta)$ in collinear states in \SCETb\ should vary as corresponding powers of  $Q\lambda^2$.  Similarly, each rapidity integral should behave as $\lambda^{2(1-a)}$. Combined with the factor $1/Q$ in front of the rapidity integral, power corrections to the collinear jet function occur as powers of $\lambda^{4-2a}/\tau_a = (1/\tau_a)(\Lqcd/Q)^{2-a}$.  Correspondingly,  in Laplace moment space, this becomes
a power series in $\nu(\Lqcd/Q)^{2-a}$.  The latter is also the only argument for the 
jet function that serves as a boundary condition in the perturbative QCD
 resummation of Ref.~\cite{BKS03}.\footnote{See, for example, Eqs.\ (67) and (74) of \cite{BKS03}.}
As long as $a<1$, we may consider these to be subleading compared to the power corrections of the soft function, which are powers of $\Lqcd/Q$. For $a\gtrsim 1$, we must take them into account, along with the recoil corrections mentioned above \cite{scaling1,scaling2,MW94}. 

From now on, we consider only observables that pick out jets with typical transverse momenta well above the nonperturbative scale. 
In the language of SCET, this allows us to work  in the theory \SCETa\ and consider power corrections only  from the soft function.

\section{Momentum Flow Operators, Universality and Scaling}

\subsection{Nonperturbative Universality from Perturbative QCD}

A striking  prediction from the analysis of event shapes
in perturbation theory, including those given
 in Eq.~(\ref{shapedefs}),
is the universality of power corrections to their mean values
\cite{KS95,KS99,BKS03,scaling1,scaling2,Salam01,DW1,DMW,DW2,Milan,flow},
\bea
\langle e \rangle =  \langle e \rangle_{\text{PT}} + c_e\frac{ \mathcal{A}}{Q}\, .
\label{avgshift}
\eea
  In this expression, ${\cal A}$ a universal parameter and $c_e$ is a
calculable coefficient that depends on the observable, as we shall see below.
The same reasoning that leads to (\ref{avgshift}), 
when applied 
to the event shape distributions, produces a shift in the
resummed perturbative cross section,
\begin{equation}
\label{shift}
\frac{d\sigma}{d e}(e)\biggr\rvert_{\text{PT}} \underset{\text{NP}}{\longrightarrow} 
\frac{d\sigma}{de}\left(e - c_e \frac{\mathcal{A}}{Q}\right)\biggr\rvert_{\text{PT}} .
\end{equation}
These relations were derived in  Refs.\ \cite{DW1,DMW,DW2} from 
the assumption of a ``dispersive" representation for
$\alpha_s(\mu^2)$ considered as an analytic function
of the scale $\mu$, and
in Refs.\  \cite{KS95}
  they were abstracted directly from the form of resummed perturbation
theory.

A more general approach \cite{KS99,scaling1,scaling2} replaces the shift 
of Eq.~(\ref{shift}) by a convolution with
a shape function defined as above, which reduces to a 
product in Laplace moment space, Eq.~(\ref{factoredLaplace}).   
As we have noted, these shape functions are all different, but
for the angularities a generalization of the universality of Eq.~(\ref{avgshift}) has been
suggested, in the form of a scaling relation.
The Laplace-transformed shape function for angularity distributions arising from 
resummed perturbation theory
at 
next-to-leading logarithm (NLL)
 \cite{scaling1,scaling2} displays a simple scaling with the parameter $a$:
\begin{eqnarray}
\label{Laplace}
\ln S_a(\nu) = \frac{1}{1-a}\sum_{n=1}^\infty\lambda_n\left(-\frac{\nu}{Q}\right)^n,
\end{eqnarray}
where $\lambda_n$ is independent of $a$.    If we keep only the linear, $\nu/Q$, term
in the shape function, its inverse Laplace transform gives a delta function,
which in the convolution of Eq.~(\ref{firstfact}) leads
immediately to the shift of Eq.~(\ref{shift}).
As noted above, we limit our attention to angularities for $a<1$.
The values of the  
coefficients $\lambda_n$ of Eq.~(\ref{Laplace}),
of course, must be abstracted from a combination of experiment and
resummed perturbation theory.

Event shape functions derived from resummation organize
all corrections in $\nu/Q$ that are implied by 
perturbation theory.  Formally,
the coefficients $\lambda_n$ 
are given in the NLL resummed cross section by
\bea
\lambda_n =
\frac{2}{n\, n!}\, {\left(-\frac{\nu}{Q}\right)}^{n}
\int\limits_{0}^{\kappa^2} \frac{dp_T^2}{p_T^2}\; p_T^n\;
A\left(\alpha_s(p_T)\right)\, ,
\label{NLLlambda}
\eea
where $A\left(\alpha_s(p_T)\right) = C_i(\alpha_s/\pi) + \dots$ is the residue of the
$1/(1-x)$ pole in the splitting function for the parton, $i=$ quark or gluon,
that initiates the jet, and $\kappa$ is an infrared
factorization scale.   In this picture of power corrections, the coefficients
$\lambda_n$ are independent of $a$.   
The coefficient of the lowest power, $n=1$, is equivalent to an integral
over the running coupling, defined in a scheme where the coupling
 incorporates all higher powers of $A(\alpha_s)$ in $\overline{\rm MS}$
\cite{CMW91}.
This approach generalizes
the dispersive treatment of Refs.\ \cite{DMW,DW2,Milan} to higher
power corrections, but shares with it a reliance on (exponentiated)
low-order gluon emission.   

Analyses based on a dispersive coupling or on
resummation rely to a greater or lesser extent on the kinematics
of single soft gluon emission in the final state, and the 
universality relations follow from the boost invariance of these emission cross
sections.  The ``Milan 
factor" \cite{Milan} of the dispersive approach
accounts for effects at next order in $\alpha_s$, where 
boost invariance and the resultant universalities can be
maintained. 
We are about to show that the boost invariance of Wilson lines in the soft shape function Eq.~(\ref{shapefunction}) 
is by itself  enough to prove the universality relation for the mean values without 
further assumptions.  In Refs.\ \cite{KS99} and \cite{flow} the role of energy flow 
was explored in a manner closely related to our discussion below.

\subsection{Universality in Average Event Shapes from the Soft Function}

We continue to limit our attention to event shapes for which 
the dominant power corrections come from the soft function.
For the shape function in the form given in Eq.~(\ref{shapewithXu}),
the operators in the matrix element no longer contain any reference to the final state $X_u$, so,
as in Eq.~(\ref{shapewithoutXu}),
 we may perform the sum over intermediate ultrasoft states, leaving
\begin{align}
S_{e}(e) = \frac{1}{N_C}\Tr\bra 0 \overline{Y}_{\bar n}^\dag Y_n^\dag \delta\left(e - \frac{1}{Q}\int\! d\eta\,\mathcal{E}_T(\eta)f_e(\eta)\right) Y_n\overline{Y}_{\bar n}\ket{0}.
\label{shapenoXu}
\end{align}
From now on we drop the explicit dependence of the soft function on the scale $\mu_s$. In (\ref{shapenoXu}), we insert factors of $U(\Lambda(\eta'))^\dag U(\Lambda(\eta')) = 1$, implementing a Lorentz boost of each operator in the $z$-direction with a rapidity $\eta'$. 
The vacuum $\ket{0}$ is invariant under Lorentz boosts, and the Wilson lines are also invariant:
\begin{align}
U(\Lambda(\eta'))Y_n(0) U(\Lambda(\eta'))^\dag & 
= U(\Lambda(\eta')) P\exp\left[ ig\int_0^\infty\! ds\, n\cdot A_{us}(ns)\right] U(\Lambda(\eta'))^\dag \\
& {=}\ P\exp\left[ig\int_0^\infty \!ds\,\alpha n\cdot A_{us}(\alpha ns)\right] = Y_n(0),\nonumber
\end{align}
where $\alpha = \e^{-\eta'}$, as $n\rightarrow \alpha n$ and $\bar n\rightarrow \alpha^{-1}\bar n$. (This is also known in SCET as type-III reparametrization invariance \cite{RPI}.) The only change is in the operator $\mathcal{E}_T(\eta)$:
\begin{equation}
\begin{split}
U(\Lambda(\eta'))\mathcal{E}_T(\eta)U(\Lambda(\eta'))^\dag =\mathcal{E}_T(\eta+\eta')\, ,
\label{Eboost}
\end{split}
\end{equation}
which follows from the defining relation for the ${\cal E}_T$ operators, Eq.~(\ref{TEF}).
Thus, the argument of the operator $\mathcal{E}_T(\eta)$ in the shape function in Eq.~(\ref{shapenoXu}) may be shifted to any value of rapidity, $\mathcal{E}_T(\eta)\rightarrow \mathcal{E}_T(\eta+\eta')$. 
At this stage, this does not yet allow us to perform the rapidity integral of $f_e(\eta)$ inside the delta function. Thus we do not find that the leading power correction simply shifts the argument of the perturbative event shape distributions, as the delta function 
is a highly nonlinear function of the energy flow operator and sits sandwiched between Wilson lines in the matrix element.  If we  do neglect correlations
between these operators,
we derive a delta function for 
the shape function, and reproduce the 
shift in the distribution, Eq.~(\ref{shift}) \cite{KS99,flow}.  

The boost property (\ref{Eboost}) of a single operator, however,  
gives a strong result when applied to
 the first moment of an event shape distribution \cite{A001}.
 Taylor expanding the delta function in Eq.~(\ref{shapenoXu}) (which is valid if we integrate the distribution over a sufficiently large region near the endpoint), we find
\begin{equation}
\begin{split}
S_{e}(e) = \delta(e) - \delta'(e) \frac{1}{Q}\int\! d\eta\, f_e(\eta)  \frac{1}{N_C}\Tr\bra{0}\overline{Y}_{\bar n}^\dag Y_n^\dag\mathcal{E}_T(\eta+\eta') Y_n\overline{Y}_{\bar n}\ket{0} + \cdots\, .
\end{split}
\end{equation}
 Recalling the boost properties of the Wilson lines and the energy flow operators 
  $\mathcal{E}_T(\eta)$, we are free to choose any value for $\eta'$
  in this expression.  Then, choosing $\eta'=-\eta$, we find that,
remarkably,  we may take the
matrix element of the $\mathcal{E}_T$ operator out of the integral over $\eta$, leaving the result
\begin{equation}
S_e(e) = \delta(e) - \delta'(e)c_e\frac{\mathcal{A}}{Q} +\cdots,
\end{equation}
where the coefficient $c_e$ is given by the integral,
\begin{equation}
c_e = \int_{-\infty}^\infty d\eta\,f_e(\eta),
\label{ccalc}
\end{equation}
and the universal quantity $\mathcal{A}$ is 
\begin{equation}
\mathcal{A} = \frac{1}{N_C}\Tr\bra{0}\overline{Y}_{\bar n}^\dag Y_n^\dag \mathcal{E}_T(0)Y_n\overline{Y}_{\bar n}\ket{0}\, .
\label{calAdef}
\end{equation}
 For the $C$-parameter and angularities $\tau_a$, the integrals of the 
 corresponding weight functions,
\begin{equation}
f_C(\eta) = \frac{3}{\cosh\eta},\quad f_{\tau_a} = \e^{-\abs{\eta}(1-a)},
\end{equation}
over all rapidities give the coefficients,
\begin{equation}
c_C = 3\pi,\quad c_{\tau_a} = \frac{2}{1-a}.
\end{equation}
When convoluted with the perturbative distribution, $S_e(e)$  reproduces the universality relations of Eq.~(\ref{avgshift}) for the first moments of the distributions. We have thus established these results without appealing to a one-gluon or related 
approximation.  
All higher-order corrections 
due to multiple-gluon emission separate from the 
observable-dependent factor $c_e$, which can be computed
in a ``naive" fashion \cite{review} as in Eq.~(\ref{ccalc}) above.

The result for the $C$-parameter may be extended to a larger class of related event shapes by defining functions, $f_{C_a}(\eta) = 3/\cosh^a \eta$, by analogy to the angularities. The integral over rapidities of this function gives the coefficient $c_{C_a} = 3B(a/2,1/2)$, where 
$B(x,y)$ is the beta function. In like manner, various new event shapes may be defined by appropriate choices for the function $f_e(\eta)$.

\subsection{Angularity Distributions and Momentum  Flow}

The expression (\ref{shapenoXu}) for the shape function in terms
of energy flow operators enables us to put 
the power expansion of Eq.\ (\ref{Laplace})
 into a more general field-theoretic context, and to discuss the 
possible significance of scale breaking.

Let us compare Eq.~(\ref{Laplace}), derived from resummed
perturbation theory, with the Laplace transform of the 
corresponding shape function in Eq.~(\ref{shapenoXu}) 
  \cite{flow}.   This is given by
\begin{equation}
\tilde S_a(\nu) = \frac{1}{N_C}\Tr \bra 0 \overline{Y}_{\bar n}^\dag Y_n^\dag
\exp\left[ - \frac{\nu}{Q}\int\! d\eta\,
\e^{-\abs{\eta}(1-a)} \mathcal{E}_T(\eta)
\right] \;Y_n\overline{Y}_{\bar n}\ket{0} ,
\label{SaLaplace}
\end{equation}
which can be re-expressed as an expansion in cumulants,
\begin{align}
\ln\left[\frac{\tilde S_a(\nu)}{\tilde S_a(0)}\right] &=
 \sum_{n=1}^\infty\frac{1}{n!}\left(-\frac{\nu}{Q}\right)^n
 \Cumulant{\left[\int d\eta\,\e^{-\abs{\eta}(1-a)}\mathcal{E}_T(\eta)\right]^n}
\nonumber\\
&\equiv  \sum_{n=1}^\infty\frac{1}{n!}\left(-\frac{\nu}{Q}\right)^n\; {\cal A}_n(a) .
\label{cumulants}
\end{align}
Here, and below, in the cumulants the Wilson lines $Y_n$ and $Y_{\bar{n}}$
are understood.  With this normalization, the coefficient
${\cal A}_1(a)$ for the angularities is related to the universal coefficient 
${\cal A}$ in Eq.~(\ref{calAdef}) by
${\cal A}_1(a) = 2{\cal A}/(1-a)$.
 The factor of $\tilde S_a(0)$ on the left-hand side 
 of Eq.~(\ref{cumulants})
 correctly accounts for the normalization of the soft function. (Of course, from Eq.~(\ref{SaLaplace}), we see that $\tilde S_a(0) = 1$, but the normalization would not be trivial in the analogous equation for the jet function, for instance.)
In terms of the matrix elements above, 
we find a general form for the coefficients $\lambda_n$, 
which is not limited to NLL resummation, 
\bea
\lambda_n(a) =
\frac{1-a}{n!}\ {\cal A}_n(a)\, ,
\label{lambdacumulant}
\eea
which, in the general case  for $n>1$, may still depend upon $a$, as indicated.

To explore the information contained in the cumulants, $\mathcal{A}_n$, let us study the $a$ dependence of the parameters $\lambda_n$ in Eq.~(\ref{lambdacumulant}) for low $n$. 
The $n=1$ term, $\lambda_1(a)$, is independent of $a$, as we showed in the previous section, in agreement with the resummed perturbation theory result, Eq.~(\ref{NLLlambda}).
The $a$ dependence of the second and higher terms, however, differs in
general.  Nevertheless, boost invariance always allows us to perform one rapidity integral
in the cumulant matrix elements.
For the case $n=2$, we have
\begin{eqnarray}
\lambda_2(a) =\frac{1}{2}\int_{-\infty}^\infty d\eta 
\Bigl[1 + (1-a)\abs{\eta}\Bigr]\e^{-\abs{\eta}(1-a)} \cumulant{\mathcal{E}_T(0) \mathcal{E}_T(\eta)}.
\label{2ndcumulant}
\end{eqnarray}
Under certain conditions, the $a$ dependence of this expression also disappears. 
In Ref.~\cite{scaling1}, it was observed that the scaling rule for the $n$th cumulant term in Eq.~(\ref{cumulants}) is good when the energy flow correlations are negligible for rapidity intervals larger than a range $\Delta\eta\sim 1/[n(1-a)]$. Assume, then, that the correlator $\cumulant{\mathcal{E}_T(0) \mathcal{E}_T(\eta)}$ is nonzero only for 
$\eta\ll \frac{1}{2(1-a)}$. Then we may Taylor expand the remainder of the integrand in Eq.~(\ref{2ndcumulant}) about $\eta = 0$:
\begin{equation}
\int_0^\infty d\eta \biggl\{1  - \frac{1}{2}[(1-a)\eta]^2 
 + \frac{1}{3}[(1-a)\eta]^3 
 + \cdots\biggr\}
 \cumulant{\mathcal{E}_T(0) \mathcal{E}_T(\eta)}.
\end{equation}
Insofar as the correlator $\cumulant{\mathcal{E}_T(0) \mathcal{E}_T(\eta)}$ has support only over a region $\eta\ll \frac{1}{1-a}$, the leading term of the expansion dominates, and we recover the $a$-independence of $\lambda_2$.  Interestingly, there is no $\mathcal{O}(\eta)$ term in the expansion multiplying the correlator, so that violations of the scaling rule should be even smaller than one might initially expect,
at least for moderate values of $1-a$.

We must wait on the analysis of data to interpret the significance of
scale breaking for the angularities.  Supposing, however, that substantial
scale breaking were found in the power 
$(\nu/Q)^2$
in the shape functions
for angularities, we can learn about nonperturbative correlations in
energy flow through Laplace moments of the cumulants.  
For example, using Eq.~(\ref{2ndcumulant}), we observe that
\bea
{\cal C}_2(a) - \frac{\partial}{\partial \ln(1-a)}\, {\cal C}_2(a)
=
 \lambda_2(a),
\label{diffeq}
\eea
where ${\cal C}_2(a)$ is a direct Laplace moment of the correlation operators
in terms of their rapidity separation, with one fixed at rapidity
zero,
\bea
 {\cal C}_2(a) \equiv 
\frac{1}{2} \int_{-\infty}^\infty d\eta\; e^{-(1-a)|\eta|}\; 
 \cumulant{\mathcal{E}_T(0) \mathcal{E}_T(\eta)}\, .
\label{calCdef}
\eea
Assuming that the correlations vanish for 
$a\rightarrow -\infty$, 
the solution to Eq.~(\ref{diffeq}) gives these Laplace moments
directly in terms of the cumulants ${\cal A}_2$, and hence in terms of the coefficient
$\lambda_2(a)$, which is, in principle, an observable,
\bea
{\cal C}_2(a) = (1-a)\; \int_{-\infty}^a \frac{da'}{(1-a')^2}\, \lambda_2(a')\, .  
\eea
Furthermore, derivatives of ${\cal C}_2(a)$ with respect to $a$
provide information on Laplace moments of the cumulants supplemented by powers.

For higher $n$, the situation becomes somewhat more complex, but continues
to encode potentially interesting physical information.  
For the coefficient
 $\mathcal{A}_3$ in Eq.~(\ref{cumulants}), we 
can similarly perform one of the three rapidity integrals and obtain
\begin{equation}
\begin{split}
\mathcal{A}_3(a)&= \frac{4}{1-a}
\int_0^\infty d\eta_2\int_0^\infty d\eta_3\, \e^{-(\eta_2+\eta_3)(1-a)} \\
&\qquad\times\left[3\cumulant{\mathcal{E}_T(0)\mathcal{E}_T(-\eta_2)\mathcal{E}_T(\eta_3)} - \cumulant{\mathcal{E}_T(0)\mathcal{E}_T(\eta_2)\mathcal{E}_T(\eta_3)} \right],
\end{split}
\end{equation}
which again respects
the $\frac{1}{1-a}$ scaling under the assumption that the correlators are nonzero only for $\eta_{2,3}\ll \frac{1}{1-a}$. 

\section{Conclusion}

We have explored power corrections for event shapes
using
 factorization theorems in both full QCD and SCET.
In this context, we have shown that the formalisms lead to equivalent
event  shape functions that summarize 
nonperturbative effects of soft gluon emission on event 
 shape distributions for two-jet events.   We have shown how the
 boost invariance of lightlike Wilson lines implies the 
 universality of the leading $1/Q$ corrections to the mean values of the event shapes,
 without relying on low-order or even resummed perturbative calculations.
 
 In addition, we have used the field-theoretic formalism to 
 interpret potential violations of
the scaling rule for angularity shape functions
in terms of correlations 
between energy flow operators
for soft gluon radiation.
Using $1/Q^2$ corrections in shape functions as an example,
we have demonstrated how, in principle, 
a violation of scaling for the angularities can provide information on 
specific matrix elements in the effective theory.  The analysis of existing 
extensive and high-quality data
from leptonic annihilation experiments could,
in this way, provide a new experimental window 
into the process of hadronization in quantum chromodynamics.

\begin{acknowledgments}

We would like to thank C.\ F.\ Berger, M.\ Kaur and
A.~Manohar for helpful discussions, and S.~D.~Ellis for 
reading the manuscript.   
CL thanks the C.~N.~Yang Institute for Theoretical Physics for hospitality during a portion of this work.
The authors would like to thank the organizers of the TASI 2004 summer school 
where conversations led to some of the questions addressed above, 
and the organizers of the 2006 FRIF workshop at the LPTHE, where 
the collaboration that led to this work was begun.
The work of CL was supported in part by the U.S. Department of Energy under grant number DE-FG02-00ER41132. The work of GS was supported in part by the National Science Foundation, grants PHY-0354776 and PHY-0345922.

\end{acknowledgments}

\end{document}